**Surface recombination and space-charge-limited photocurrent-voltage (PC-V) measurements in (Cd,Mn)Te samples. Kinetics of photocurrent (PC).**

Andrzej Mycielski[1,*], Dominika M. Kochanowska[1], Aneta Wardak[1], Krzysztof Gościński[1], Michał Szot[1,2], Witold Dobrowolski[1], Gabriela Janusz[1], Małgorzata Górska[1], Łukasz Janiak[3], Wiesław Czarnacki[3], Łukasz Świderski[3], Joanna Iwanowska-Hanke[3] and Marek Moszyński[3]

[1] Institute of Physics, Polish Academy of Sciences, Aleja Lotników 32/46, PL-02668 Warsaw, Poland;

dmkoch@ifpan.edu.pl (D.M.K.); wardak@ifpan.edu.pl (A.W.); kgosc@ifpan.edu.pl (K.G.); szot@ifpan.edu.pl (M.S.); dobro@ifpan.edu.pl (W.D.); gjanusz@ifpan.edu.pl (G.J.); gorska@ifpan.edu.pl (M.G.);

[2] International Research Centre MagTop, Institute of Physics, Polish Academy of Sciences, Aleja Lotników 32/46, PL-02668 Warsaw, Poland;

[3] National Centre for Nuclear Research, Andrzeja Sołtana 7, PL-05400 Otwock, Poland;

lukasz.janiak@ncbj.gov.pl (Ł.J.); wieslaw.czarnacki@ncbj.gov.pl (W.C.); lukasz.swiderski@ncbj.gov.pl (Ł.Ś.); joanna.iwanowska@ncbj.gov.pl (J.I.-H.); marek.moszynski@ncbj.gov.pl (M.M.)

[*] Correspondence: mycie@ifpan.edu.pl

**Abstract**

Photocurrent-voltage characteristic (PC-V) is a method of determining the critical parameter in X-ray and gamma-ray detector plates, i.e., the carrier mobility – lifetime product, $\mu\tau$. We show on the (Cd,Mn)Te samples that the measurement results depend strongly on the surface treatment and the charge space distribution. The PC-V characteristics obtained for $\hbar\omega > E_g$ and $\hbar\omega \sim E_g$ indicated that etching with 20% HCl caused an appearance of a significant concentration of very shallow surface traps at the (Cd,Mn)Te sample surface. These traps seriously changed the measurements of PC-V characteristics and PC kinetics. We also noticed a small contribution of holes to photoconductivity in the PC kinetics. The PC-V characteristics measurements for $\hbar\omega > E_g$ may test the detector plate surface quality.

1. **Introduction**

(Cd,Mn)Te is a wide bandgap semiconducting compound. We want to make it competitive with commonly used CdTe and (Cd,Zn)Te compounds for X-ray and gamma-ray detectors. The attained values of the resistivity, $\rho \sim 10^{10}$ $\Omega$cm, and the $\mu\tau$ product, $\mu\tau > 10^{-3}$ cm$^2$V$^{-1}$, make us hope for the commercial use of the material. A review of (Cd,Mn)Te studies as material for those detectors may be found in Ref. [1]. One of the critical parameters is the $\mu_e\tau_e$ product, where $\mu_e$ is the carrier mobility (here electrons) and $\tau_e$ is the carrier lifetime (here also electrons). It is not easy to attain a large $\mu_e\tau_e$ parameter ($> 5\times10^{-3}$ cm$^2$V$^{-1}$), uniform in the whole volume of the sample with the area of 20×20 mm$^2$ or more. Usually, the $\mu_e\tau_e$ parameter is determined by illuminating the sample with alpha-rays ($\alpha$) or gamma-rays ($\gamma$) and investigating the charge collection efficiency (CCE). CCE is defined as the fraction of injected charges that is collected in the external circuit. For alpha-rays with an energy of 5.5 MeV, the stopping distance is about 20 µm. It is like a surface effect compared with the typical sample thickness (~ 2000 µm). The penetration is much deeper for the gamma radiation (Am-241 59.6 keV). The attenuation length is about 180 µm, it is ~ 0.1 of the sample thickness. The CCE value may be obtained by fitting the experimental PC-V data with the Hecht-Many function [2,3], where the fitting parameters are $\mu_e\tau_e$ and the surface recombination velocity, $s_e$.

The Hecht-Many model assumes no space charge in the sample in stationary conditions. Therefore, the internal electric field is uniform, $E = VL^{-1}$, where $E$ is the field, $V$ is the applied voltage, and $L$ is the sample thickness. The model also assumes that the parameters $\mu_e$ and $\tau_e$, are uniform in the sample volume. For (Cd,Zn)Te Cui *et al.* [4,5] have shown that the $\mu_e\tau_e$ product may be determined by measuring the photocurrent as a function of the applied voltage (PC-V). The energy of the light used for the electron injection was higher or smaller than the energy gap, $E_g$. The experiments indicated that the results strongly depended on the surface conditions and the surface processing methods. The authors recall similar observations of the impact of surface conditions on the carrier transport measurements in HgI$_2$ [6–8].

This study aimed to perform comparable measurements for (Cd,Mn)Te using the PC-V technique and different methods of preparing the sample surface. Similar experimental and theoretical studies were performed for (Cd,Zn)Te [9–13]. Shen *et al.* [14] showed the interesting results of processing the (Cd,Mn)Te surfaces. The main conclusion of that study was that the final etching with 5% HCl created the smoothest surface with the average value of the surface roughness, $R_a$, 0.84 nm. We decided to test Shen's method. We etched the polished surfaces of the (Cd,Mn)Te samples with HCl, changing the concentration of the solution.

Below we present results of measurements of the PC-V characteristics. The results were compared with the measurements using the γ-rays (Am-241 59.6 keV). We show that the results of the PC-V investigation are significantly different for different methods of the sample surface preparation. Finally, we present the PC-V characteristics for the excitation by the light with energy above $E_g$ and slightly below $E_g$.

2. **Materials and Methods**

The crystals of $Cd_{0.95}Mn_{0.05}Te:In$ ($[In]\approx1\times10^{17}$ cm$^{-3}$) were prepared by the low-pressure Bridgman (LPB) method in the vertical configuration. To obtain semi-insulating materials with $\rho \sim 10^{10}$ Ωcm the crystals were doped with indium. To obtain deep donors $Te_{Cd}^{2+}$ [15,16], tellurium was added in proportion 50-100 mg Te per 100 g of material. The crystallization speed was 0.5 – 1.0 mmh$^{-1}$. Before cutting, the crystals were etched with the K. Durose etchant [17] to reveal twins and grain boundaries. Then, the samples were cut parallel to the twin planes (111). The samples were initially ground and etched with K. Durose etchant to determine the planes of A (cadmium) and B (tellurium). After that, the samples were mechanically ground using a water suspension of 4.5 μm SiC powder, and mechano-chemically polished in 2% bromine-methanol solution.

Min Shen *et al.* [14] described interesting methods of preparing surfaces of (Cd,Mn)Te samples for depositing contacts. The results inspired us to use HCl in the procedure of the sample surface preparation. The last stage of the authors' method is chemical polishing (CP) in 5% HCl. We prepared three sets of samples, the surfaces of which were treated with different-concentrated HCl, namely: 10%, 15%, or 20% for 10-20 s.

Onto both types of surfaces thin contact layers Au/Pd or Pt/Pd with thicknesses ~ 30 nm were deposited using photolithography or the sputtering method. Palladium added to Au or Pt target increases adhesion. The monocrystalline samples had dimensions of 10×10 mm$^2$ or 20×20 mm$^2$ and thicknesses of 1–3 mm. In the center of the sample surface, the metal contact layers (anode and cathode) had diameters $\varphi = 5$ mm for samples with dimensions 10×10 mm$^2$ and $\varphi = 8$ mm for samples with dimensions 20×20 mm$^2$. On the surfaces close to the sample edges, rings, whose role was to insulate the current between anode and cathode from the leakage current on the sidewalls of the sample, were deposited by sputtering or photolithography. For PC-V measurements, the central part of the cathode was illuminated through an aperture with $\varphi = 3$ mm for samples with dimensions 10×10 mm$^2$ and $\varphi = 5$ mm for samples with dimensions 20×20 mm$^2$. The light sources were light-emitting diodes (LEDs), and the light passed through interference filters (IF) to reduce the width of a spectral range of the exciting radiation. The incident light power in μW, converted to eV/s and divided by the photon energy, enabled us to determine the number of incident photons.

Each photon, reaching the sample surface through the metal contact, especially with the energy $\hbar\omega > E_g$ creates an electron-hole pair. For $\hbar\omega$ much greater than $E_g$ the absorption coefficient in our samples is about $2\times10^4$ cm$^{-1}$ [18,19]. The radiation is absorbed at a distance from the surface smaller than 1 μm. Holes quickly return to the cathode, and electrons conduct the electric current.

So, the current measured in the PC-V characteristic is an electron current. The PC-V characteristics were measured at a constant illumination of the middle part of the cathode by increasing the bias voltage. Kinetic of photocurrent was measured at a selected constant bias voltage by applying a nearly rectangular light pulse (pulse rise time ~ 0.5 ms). We excited the samples with photon energies $\hbar\omega > E_g$ and $\hbar\omega \sim E_g$.

The resistivity maps of all investigated samples were obtained by a contactless method – Time-Dependent Charge Measurement (TDCM) method with the use of EU-$\rho$-$\mu\tau$-SCAN apparatus.

The experiments were performed with the use of the following equipment:

- Two Keithley 6517B Electrometers were used to apply the bias voltage and to measure the current between cathode and anode and the current between rings on the cathode and anode sides. The rings, the cathode, and the anode were biased with the same voltage.
- Illuminating LEDs were powered by the Keithley 6220 Precision Current Source.
- The LED light passing through IF was uncovered and covered by the Thorlabs SC10 Shutter Controller. Radiation power was measured by the Thorlabs PM100D Optical Power Meter.
- PC kinetic was registered by the Tektronix MSO 54 Oscilloscope by using the pre-amplifier from the Keithley 6517B Electrometer.
- The PC-V and I-V characteristics were measured at a temperature of 306 K (33°C), higher than room temperature 294 K (21°C), using a slightly heated sample holder.

**3. Results and Discussion**

**3.1. Photocurrent-voltage characteristic**

In the first set of Cd$_{0.95}$Mn$_{0.05}$Te:In samples, the sample surfaces were etched with 10% or 15% HCl. The PC-V characteristic of one of the samples is shown in Figure 1. The experimental points were fitted with the Hecht formula:

$$J = \frac{J_0 \mu_e \tau_e E}{L}\left[1 - exp\left(-\frac{L}{\mu_e \tau_e E}\right)\right], \qquad (1)$$

where $J_0$ is the saturation photocurrent, $\mu_e$ is the electron mobility, $\tau_e$ is the electron lifetime, $E$ is the electric field intensity, and $L$ is the sample thickness. The fitting parameters were $\mu_e \tau_e$ and $J_0$. The value $\mu_e \tau_e$ obtained from the fit was about $0.81 \times 10^{-3}$ cm$^2$V$^{-1}$, and $J_0 = 3.9 \times 10^{-8}$ A.

The energy of the illuminating LED light was slightly above the energy gap, $\hbar\omega = 1.590$ eV ($E_g$ + 15 meV). The absorption coefficient for this light energy is $\alpha > 10^4$ cm$^{-1}$, so the light is absorbed in a distance from the sample surface smaller than 1 μm. For the same sample, we measured the gamma-ray induced current using 59.6 keV gamma rays from an Am-241 source. The results fitted with the Hecht formula are shown in Figure 2. The fitted curve has the same shape as the curve obtained from the PC-V measurement, and the fitting parameters are $\mu_e \tau_e = 0.76 \times 10^{-3}$ cm$^2$V$^{-1}$, $J_0 = 0.49 \times 10^{-9}$ A.

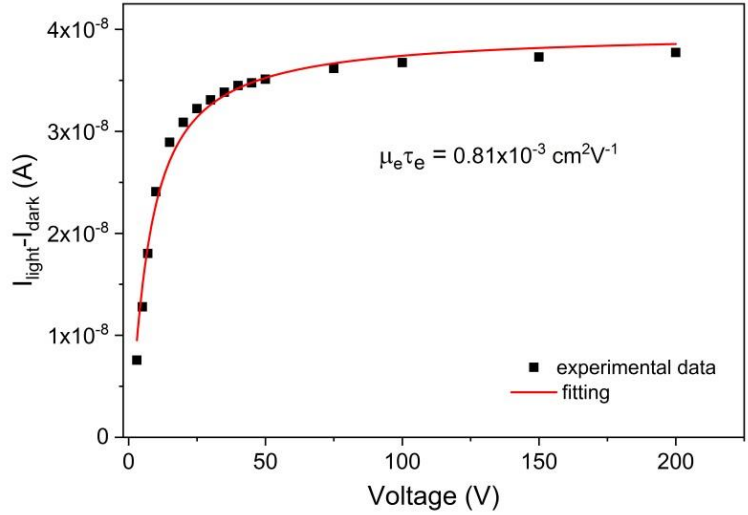

**Figure 1.** Room-temperature electron photocurrent of the Cd$_{0.95}$Mn$_{0.05}$Te:In sample as a function of a bias voltage. The solid line is a fit with the Hecht equation. The value of the $\mu_e \tau_e$ parameter is ~$0.81 \times 10^{-3}$ cm$^2$V$^{-1}$. The sample surface was prepared by etching with 10% HCl. The contacts were sputtered Au/Pd layers with thicknesses ~ 30 nm. The illuminating light energy was $\hbar\omega = 1.590$ eV ($E_g$ + 15 meV).

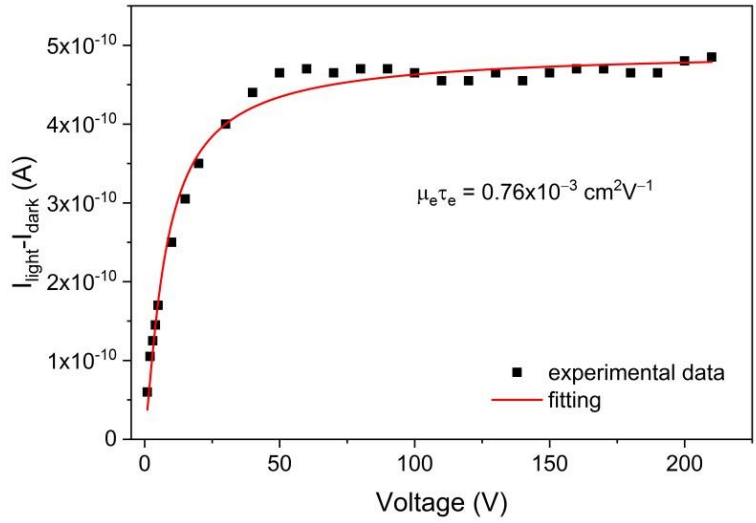

**Figure 2.** Room-temperature Am-241 gamma-ray response current for the (Cd,Mn)Te:In sample measured in Figure 1. Gamma radiation energy is 59.6 keV. The solid line is a fit with the Hecht equation, $\mu_e\tau_e \cong 0.76\times10^{-3}$ cm$^2$V$^{-1}$.

In Refs. [4] and [5], the authors observed a considerable difference in the PC-V characteristics measured on A and B surfaces of the investigated samples. We also measured PC-V characteristics of both surfaces of our samples. The results are shown in Figure 3. Results for both surfaces were fitted with the Hecht formula, and the fitting parameters were slightly different. For A surface we obtained $\mu_e\tau_e = 5.85\times10^{-4}$ cm$^2$V$^{-1}$, $J_0 = 6.9\times10^{-7}$ A, and for B $\mu_e\tau_e = 4.62\times10^{-4}$ cm$^2$V$^{-1}$, $J_0 = 7.1\times10^{-7}$ A.

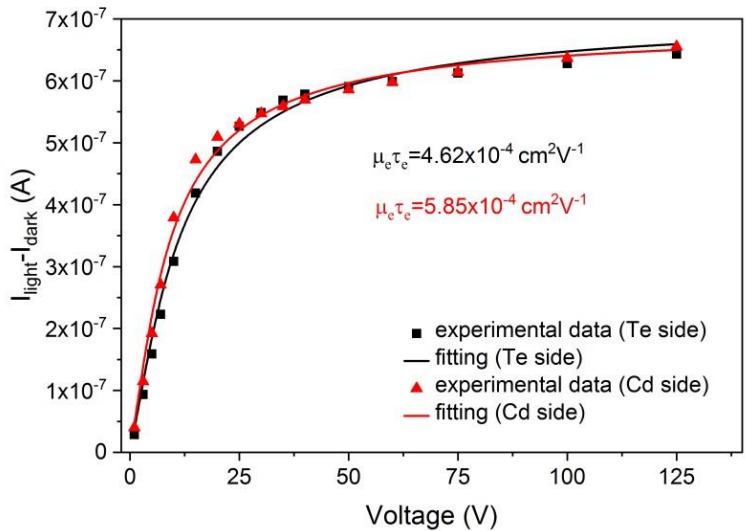

**Figure 3.** PC-V characteristic of a (Cd,Mn)Te:In sample, for two opposite surfaces (A and B). Red and black symbols and curves represent results and fittings obtained for the Cd side (A) and Te side (B), respectively. The fitting was made with the Hecht equation. Both surfaces, A and B, were etched with 15% HCl. The contacts on both surfaces of the sample were sputtered Au/Pd layers with thicknesses ~ 30 nm.

Comparing the results shown in Figures 1–3, we concluded that when the sample surfaces were treated by less-concentrated HCl, namely 10% or 15%, the results of both experiments, i.e., photocurrent measurements with the use of LED and Am-241 radiation sources, can be well described by the Hecht formula. Below we will discuss certain shortcomings of this method.

Before the PC-V characteristics, resistivity maps of all investigated samples were obtained. A typical result is shown in Figure 4. Let us pay attention to the upper left corner of the map, a sample border (and the crystal border). We see that the leakage side current may influence the measurement result up to about 3 mm from the side. Therefore, we used rings close to the sample sides in measurements of PC-V and I-V characteristics.

Semi-insulating samples, chosen for measurements of PC-V characteristics and PC kinetics, had resistivities with average values from $5\times10^9$ Ωcm to $2\times10^{10}$ Ωcm.

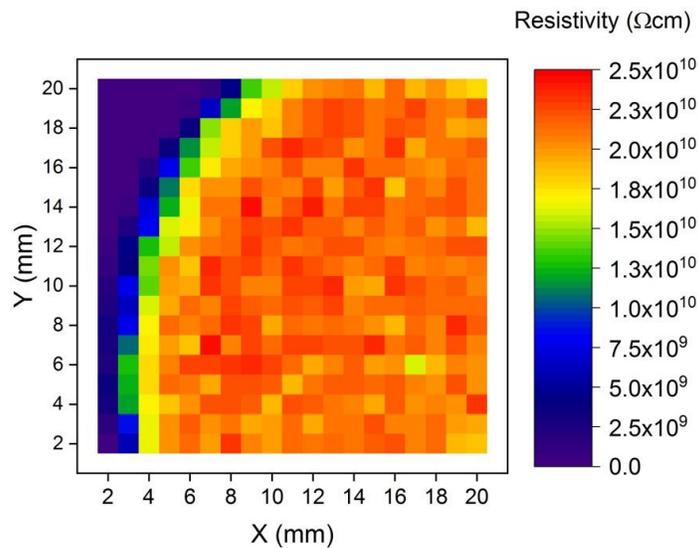

**Figure 4.** Resistivity map of a (Cd,Mn)Te:In sample measured by EU-$\rho$-$\mu\tau$-SCAN apparatus. The upper right corner is the border of the sample (and of the crystal). The side surface leakage current influences the measured sample resistivity up to about 3 mm from the side.

As was mentioned in Chapter 2, the samples were cut out of the crystals parallel to the (111) plane. Contact layers, deposited onto polished and etched surfaces, had thicknesses ~ 30 nm. In the I-V measurements, we noticed that when the anode (+) was on the A (cadmium) side, the resistivity had a higher value than when the current was applied in the opposite direction, especially for the high voltage. A typical I-V characteristic is shown in Figure 5. The right side of the plot shows the case when the anode is on the A (cadmium) side. This effect was checked on many samples with surfaces prepared by different methods. Experimentalists may pay attention to that.

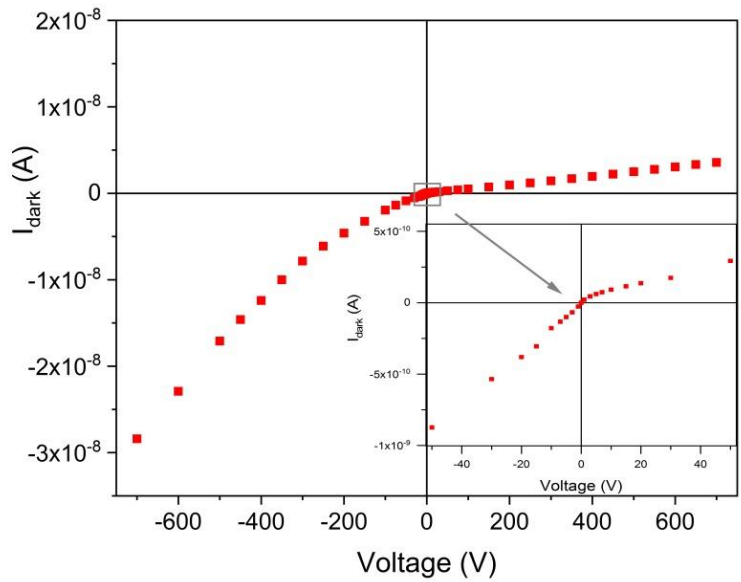

**Figure 5.** Dark I-V characteristic of a (Cd,Mn)Te**:**In sample cut parallel to the (111) plane. After preliminary preparation, the sample was etched with 20% HCl. The contacts were sputtered Pt/Pd layers. The resistivity is much higher when the anode is on the cadmium surface (A), especially at high bias voltage.

PC-V characteristics measured for a sample treated with 20% HCl are shown in Figure 6. Let us look at the first part of the PC-V curve in the range 1–75 V. It is not concave, like in the Hecht or Hecht-Many exponential formula, but slightly convex. The sample was illuminated by photons with energies of 1.590 eV, slightly above the $E_g$. This result caused us to start a series of investigations on (Cd,Mn)Te samples with surfaces prepared by etching with 20% HCl. We performed experiments with the use of interference filters, in the energy range $\hbar\omega > E_g$ and $\hbar\omega \sim E_g$, and for different light intensities. Those results are presented in Figures 6, 8, 9, and 10.

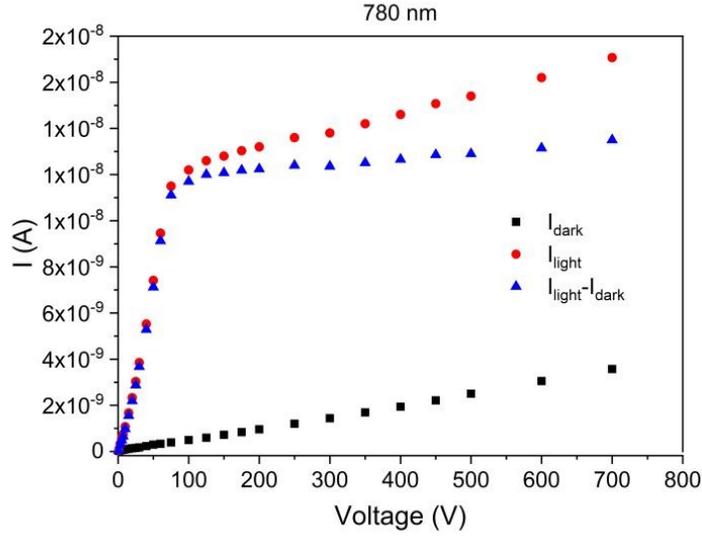

**Figure 6.** PC-V characteristic for the sample measured in Figure 5. The surface was etched with 20% HCl. At voltages up to ~ 75 V, the curve does not look like the exponential convex Hecht-Many function rather like a straight or concave line. The illumination photons energy was very close to $E_g$.

In Figure 7, we show the tail of the absorption edge for one of the $Cd_{0.95}Mn_{0.05}Te$**:**In sample. Here are also marked energies, wavelengths, and the bandwidths of the light passing through the interference filters. For example, for the interference filter 797 nm (1.556 eV), for which the absorption coefficient is $\alpha \sim 200$ cm$^{-1}$, about 90% of the radiation is absorbed in a distance of ~100 μm. To interpret the surprising results for the sample etched with 20% HCl, shown in Figure 6, we measured the PC-V characteristics using the photon energy $\hbar\omega = 1.651$ eV ($E_g$ + 76 meV), for which the absorption coefficient $\alpha \sim 2\times10^4$ cm$^{-1}$. By exciting the sample with photons with such energy, one generates charge carriers at a distance smaller than 1 μm from the sample surface. The PC-V characteristic measured at the light intensity $I_L^* = 8.49\times10^{12}$ cm$^{-2}$s$^{-1}$ is shown in Figure 8. With increasing voltage, the current increases at first as a power function, then reaches saturation and does not change up to the highest applied voltage. The experimental results shown in Figure 8 could not be fitted with the Hecht formula, and even less with the Hecht-Many formula:

$$J = \frac{J_0}{1 + \frac{s_e}{\mu_e E}} \frac{\mu_e \tau_e E}{L} \left[1 - exp\left(-\frac{L}{\mu_e \tau_e E}\right)\right], \qquad (2)$$

where $s_e$ is the surface recombination velocity.

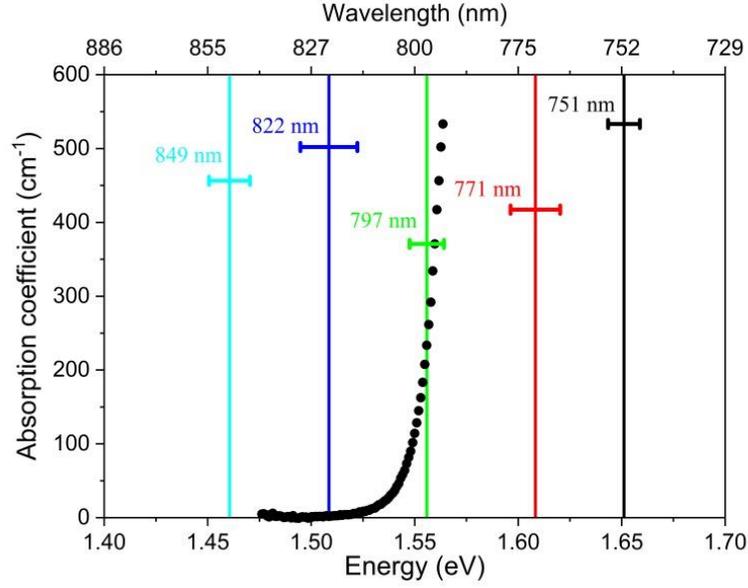

**Figure 7.** The tail of the absorption edge for a $Cd_{0.95}Mn_{0.05}Te$:In sample as a function of light wavelength used for obtaining the PC-V characteristics. The light was emitted by LEDs and passed through interference filters (IFs). Half widths of some filters are marked.

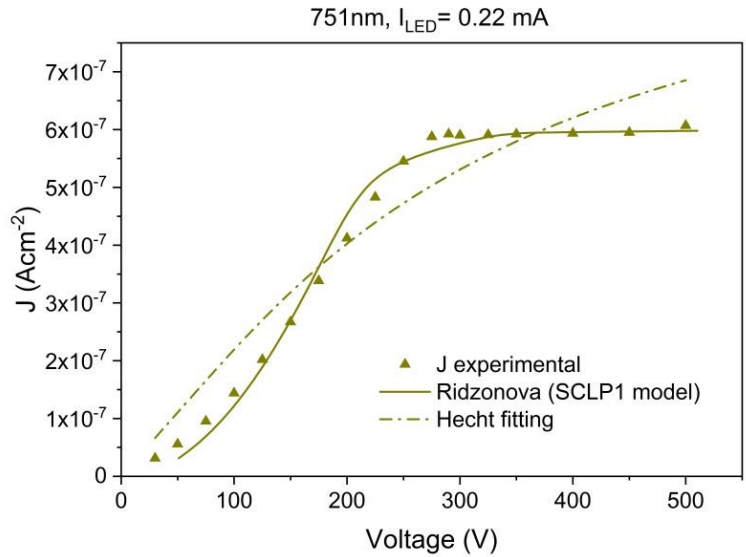

**Figure 8.** PC-V characteristic measured at the light intensity $I_L^* = 8.49 \times 10^{12}$ cm$^{-2}$s$^{-1}$. Dashed line is a fit with the Hecht formula, solid line is a fit with the Ridzonova *et al.* model [13].

We measured the PC-V characteristics at different light intensities for the same sample. The results are shown in Figure 9. For the lowest light intensity, $I_L^* = 1.39 \times 10^{12}$ cm$^{-2}$s$^{-1}$, the photocurrent saturation is reached at 150 V. For the highest light intensity, $I_L^* = 2.36 \times 10^{13}$ cm$^{-2}$s$^{-1}$, the saturation is reached at 450 V.

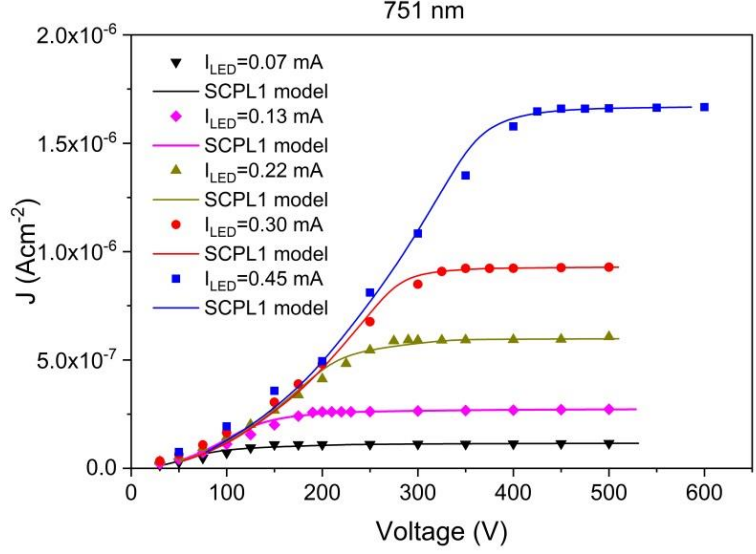

**Figure 9.** PC-V characteristics were measured at different light intensities for the sample from Figure 8. For the lowest light intensity, $I_L^* = 1.39\times10^{12}$ cm$^{-2}$s$^{-1}$, the saturation is reached at 150 V. For the highest light intensity, $I_L^* = 2.36\times10^{13}$ cm$^{-2}$s$^{-1}$, the saturation is reached at 450 V. Solid lines are fitted to the Ridzonova *et al.* model. The fitting parameters are listed in Table 1.

**Table 1.** Parameters obtained from the analysis of the PC-V data for different light intensities according to the Ridzonova *et al.* model [13]: $\theta$ and $s_e$ are fitting parameters, $N_e$ was determined from the saturation current, and $\mu_e\tau_e$ was estimated for the lowest light intensity (see text).

| LED current [mA] | $I_L^*$ [cm$^{-2}$s$^{-1}$] measured | $\theta$ | $s_e$ [cm$^1$s$^{-1}$] | $I_L$ [cm$^{-2}$s$^{-1}$] from fitting | $N_e$ [cm$^{-3}$] | $\mu_e\tau_e$ [cm$^2$V$^{-1}$] |
|---|---|---|---|---|---|---|
| 0.07 | 1.39×10$^{12}$ | 5.24×10$^{-4}$ | 4.933×10$^4$ | 7.48×10$^{11}$ | 1.7×10$^6$ | 7×10$^{-4}$ |
| 0.13 | 3.76×10$^{12}$ | 5.25×10$^{-4}$ | 3.685×10$^4$ | 1.75×10$^{12}$ | 3.1×10$^6$ | - |
| 0.22 | 8.49×10$^{12}$ | 5.08×10$^{-4}$ | 4.793×10$^4$ | 3.94×10$^{12}$ | 5.0×10$^6$ | - |
| 0.30 | 1.34×10$^{13}$ | 4.41×10$^{-4}$ | 8.280×10$^3$ | 5.84×10$^{12}$ | 6.5×10$^6$ | - |
| 0.45 | 2.36×10$^{13}$ | 4.67×10$^{-4}$ | 1.038×10$^4$ | 1.05×10$^{13}$ | 8.6×10$^6$ | - |

To interpret the results shown in Figure 9, we followed the model proposed by Ridzonova *et al.* [13] (Equation 9 there):

$$V = \frac{\varepsilon_0 \varepsilon_r \mu_e \theta}{3j} \left\{ \left[ \left( \frac{s_e j}{\mu_e (I_L e - j)} \right)^2 + \frac{2jL}{\varepsilon_0 \varepsilon_r \mu_e \theta} \right]^{3/2} - \left( \frac{s_e j}{\mu_e (I_L e - j)} \right)^3 \right\}, \qquad (3)$$

where $j$ represents the current density, $\varepsilon_0$, $\varepsilon_r$, $V$, $I_L$, $s_e$, and $e$ are the vacuum and the relative permittivity, the applied bias voltage, the light intensity which reaches the sample passing the metal layer, surface recombination velocity, and the elementary charge, respectively. Parameter $\theta$ is defined as follows:

$$\theta = \frac{n}{n + n_t}, \qquad (4)$$

where $n$ is the free electron density, and $n_t$ is the trapped electron density. Usually $n_t \gg n$, so $\theta = \frac{n}{n_t}$. In this model, the fitting parameters were $\theta$ and $s_e$. $I_L$ value was determined from the saturation of photocurrent versus voltage [13]. The model considers surface effects and the non-uniform electric field distribution in the sample, especially in the vicinity of the cathode. Fits of the experimental data with the model from Ref. [13] are shown in Figure 9.

We can see that for the sample etched with 20% HCl, the photocurrent measured at different light intensities changes exponentially with applied bias voltage, with the exponent close to *3/2*, which is shown in Figure 10. Such behavior may be expected for traps generating a "Coulomb – attractive" potential. This effect is similar to the Poole-Frenkel model [20,21] analyzed in Ref. [22]. In this paper, the de-trapping mechanism is described by the relation $E^{3/2}$, following the change of the capture cross-section proportional to $E^{-3/2}$. Such a relationship was obtained in Ref. [23] "for the decrease if the trapping cross-section as a function of electric field for germanium detectors". The results and their interpretation indicate quite deep traps.

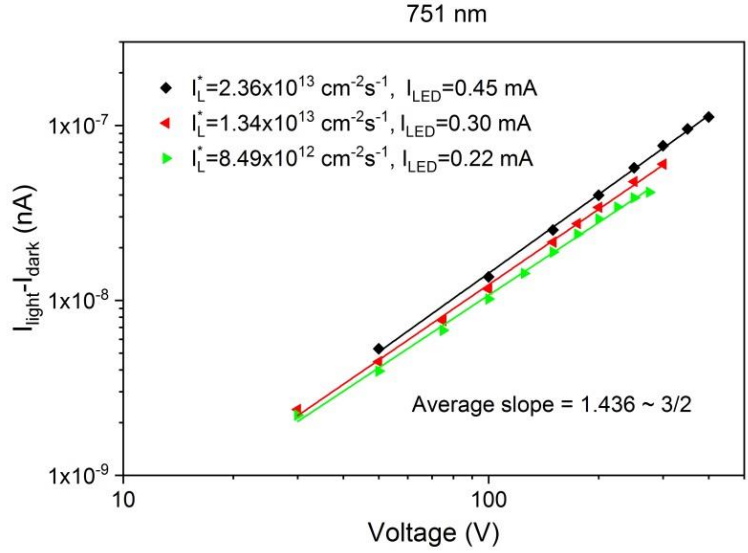

**Figure 10.** PC-V characteristics in logarithmic scale. Photocurrent increased with the bias voltage roughly like $V^{3/2}$. However, when the current is inversely proportional to the cross-section for Coulomb trapping, it should depend on the electric field like $E^{3/2}$ [20,21].

In order to explain the results presented in Figures 6, 8, and 9 we measured PC-V characteristics for another (Cd,Mn)Te sample at two slightly different temperatures: 294 K and 306 K. The result is shown in Figure 11. The small temperature change (~ 4%) caused an evident change in the measured PC-V curve. The dependence changed from a power function to a nearly exponential function at the low voltage, resembling the Hecht-Many function. Because of the presented results, we believe that we observe very shallow traps capturing electrons moving from the cathode, where they are generated by light. The increasing voltage causes de-trapping, and at the higher voltage the PC-V curve saturates. We suppose that the traps are very shallow, and their ionization at low voltage may be similar to the Poole-Frenkel effect. Moreover, for (Cd,Mn)Te crystals, the measurements of Pockels effect and Pockels E–field profiling [24–26] indicated at the high voltage the non-uniform distribution of the electric field in the sample up to about 200 μm from the cathode. Therefore, we cannot consider the electric field, $E$, inside the sample because the field is not uniform. We can take into consideration only the applied voltage.

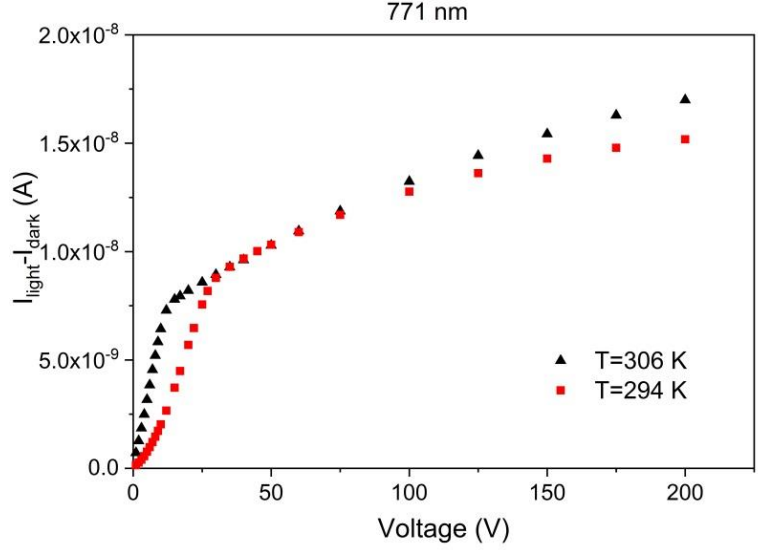

**Figure 11.** PC-V characteristics were measured at two slightly different temperatures. Below 30 V at 294 K, the dependence is power-like. It changes shape after raising temperature up to 306 K (by about 4%) and resembles the Hecht-Many function. The conducting electrons may appear due to thermal ionization of very shallow surface trapping levels.

We used the relation:

$$j = e\, N_e\, \mu_e \frac{V}{L}, \qquad (5)$$

where $N_e$ is the free electron concentration. Here we assume $\mu_e = 800\ \mathrm{cm^2 V^{-1} s^{-1}}$ [27–29]. From this relation and the saturation current, we obtained the free electron concentration, $N_e$, given in Table 1.

In Figure 9, we see for each PC-V curve a definite knee, above which the current saturates. Therefore, we believe that all electrons generated by light reach the anode for the bias voltage, $V$, corresponding to the knee. That would mean that the mean drift time, $\tau_{dr}$, corresponds to the electron lifetime, $\tau_e$ ($\tau_{dr} \approx \tau_e$).

We are aware of the possible non-uniform electric field distribution, especially for high bias voltage and high light intensity. However, we try to estimate the $\mu_e \tau_e$ product from the relation:

$$v_{dr} = \mu_e\, E = \frac{L}{\tau_{dr}}, \qquad (6)$$

Then, $L \cong \mu_e\, \tau_e\, E$ and $\mu_e\, \tau_e \cong \frac{L^2}{V}$.

We estimated the $\mu_e \tau_e$ value for the lowest light intensity as ~ $7 \times 10^{-4}$ cm$^2$V$^{-1}$. For the highest light intensity in Figure 9, the knee is at the highest voltage, and the $\mu_e \tau_e$ value is three times smaller. We realize that the electric field distribution is not uniform for higher light intensities and bias voltages, as was observed by the Pockels effect [24–26], and the estimated $\mu_e \tau_e$ is several times smaller.

### 3.2. Kinetic of photocurrent

The kinetic of photocurrent was measured on the sample, the PC-V characteristics of which were presented in Figures 8 and 9. A bias voltage was applied to the sample and a dark current stabilized after some time. Then a shutter was opened, and a nearly rectangular light illumination was applied on the cathode side. The open shutter time was about 0.5 ms. The light passed through an interference filter. The light energies were $\hbar\omega > E_g$ and $\hbar\omega \sim E_g$. Photocurrent was registered as a function of time. Then, a higher bias voltage was applied and the measurement was repeated.

The same bias voltage was applied to the electrodes and to the rings at the sample edges for every run. The current was measured with the Keithley 6517B electrometer, displayed on the Tektronix MSO 54 Oscilloscope, and registered. The time base was 180 s. Using stable light with energy 1.608 eV, which is ~ 33 meV above the energy gap and which enters the sample at a distance smaller than 1 μm, we apply consecutively higher bias voltages to obtain the results shown in Figure 12. The sharp peak at $t \sim 0$ in Figure 12 is probably an artifact from electronics. At the lowest voltage (50 V), the signal rise time till stabilization is about 40-50 s. The signal amplitude at 50 V is about ten times smaller than that one at the highest voltage (500 V). At the highest voltage, the signal rise and stabilization time reduce to ~ 3 s.

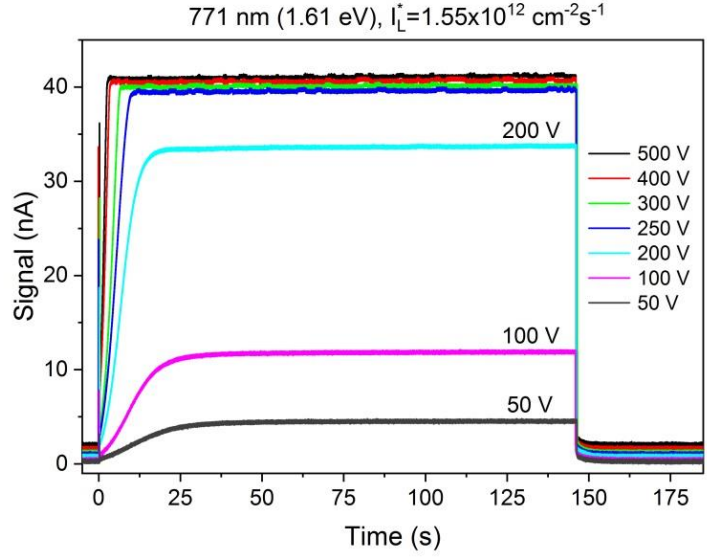

**Figure 12.** PC kinetics after rapid (~ 0.5 ms) turning on the illuminating light, $I_L^* = 1.55\times10^{12}$ cm$^{-2}$s$^{-1}$, $\hbar\omega = $ 1.608 eV ($E_g$ + 33 meV), at different voltages. Charge carriers are generated at the surface of the sample. The PC time dependence does not change much above 250 V.

The saturation current values increase with bias voltage like in the spectra shown in Figure 9. However, after shutting (close shutter time ~ 0.5 ms) the photocurrent decay is fast and does not depend on the bias voltage. Decay time has two components: (1) very short time, much less than 1 s; (2) time about 5–8 s. The time decay was not analyzed in detail in the present work.

The kinetic of photocurrent for the photon energy $\hbar\omega$ = 1.531 meV ($E_g$ − 44 meV) is much different from the one observed when $\hbar\omega > E_g$ (Figure 12), and it is shown in Figure 13. For such photon energy in the tail of the absorption edge, the absorption coefficient is about 20 cm$^{-1}$. The corresponding attenuation length is about 350 μm, about 10% of the thickness of our sample (3210 μm). The attenuation length is close to that for gamma-rays Am-241 59.6 keV, which for CdTe is about 200 μm. In Figure 13, we can see that after opening the shutter, the signal rise time at the lowest voltage (5 V) is about 2 s, and at the highest voltage (300 V) is about 1 s. Decay time, like for $\hbar\omega > E_g$, is at first very fast and after that several seconds and does not depend on the bias voltage. For $\hbar\omega$ = 1.531 eV the saturation current values increase with bias voltage like in the Hecht-Many formula. When we compare the results presented in Figures 12 and 13, taking into account the penetration depth of the radiation, we come to two important conclusions.

First, we may estimate a possible influence of holes flowing back to the cathode on the kinetic of photocurrent. For photons with energy below $E_g$, the PC kinetic is very fast, on the order of single seconds, and the hole contribution to the electric current is nearly negligible. However, at high bias voltages, $V \geq 100$ V, the signal after the first rapid rise (~1 s) increases slightly till about 100 s. This effect we attribute to the contribution of holes to the electric current.

Second, we can see that the PC kinetic strongly depends on the surface states and surface charge. The PC-V characteristics at lower voltages and the PC kinetics for $\hbar\omega > E_g$ provide a method to check for the presence of traps and a non-uniform distribution of the electric field in the detector plate volume.

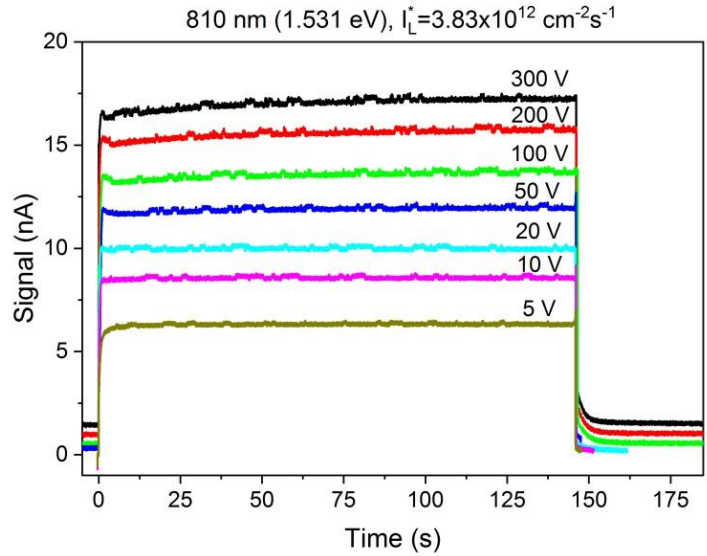

**Figure 13.** PC kinetics for the sample shown in Figure 12, $I_L^* = 3.83 \times 10^{12}$ cm$^{-2}$s$^{-1}$, $\hbar\omega = 1.531$ eV ($E_g - 44$ meV). Charge carriers are generated inside the sample, not only at the surface. The shape of the PC time dependence does not depend on the bias voltage; only the amplitude depends. The slight rise of the PC signal with time at $V \geq 100$ V was attributed to the influence of holes.

4. **Conclusions**

At present, the photocurrent-voltage characteristics are increasingly used to determine the $\mu_e\tau_e$, which is an essential parameter for X-ray and gamma-ray detectors. The present paper shows PC-V characteristics of semi-insulating (Cd,Mn)Te samples with differently prepared surfaces. We show that the different preparation of the surface leads to very different results, and in some cases, the Hecht-Many model cannot be applied. Furthermore, the PC-V characteristic features significantly changed with the concentration of HCl used after the polishing process.

We applied the final etching with 20% HCl for some of our samples, measured the PC-V characteristics, and fitted our results with the formula given by Ridzonova *et al.* [13]. The model considers surface states and space charges in the sample volume. We noticed that in the voltage region below saturation, the current is proportional to $V^{3/2}$ (in the case of uniform $E$ that would correspond to $E^{3/2}$). This is in agreement with the Poole-Frenkel model [20,21]. The $I \sim V^{3/2}$ dependence is observed even at voltages as low as about 30 V (for our samples with thicknesses ~ 0.3 cm, that would correspond to $E \sim 100$ Vcm$^{-1}$). That indicates that the electron traps formed at the surface due to the etching with highly concentrated HCl are very shallow.

We also noticed that a change in the sample temperature by 4% significantly changed the PC-V characteristic. That confirms our conclusion that the traps at the surface are very shallow.

We measured the PC kinetics for samples with a high density of shallow surface electron traps. The results obtained by excitation with phonon energy $\hbar\omega = 1.608$ eV ($E_g + 33$ meV) differ from those obtained by exciting with $\hbar\omega = 1.531$ eV ($E_g - 44$ meV). Our results show that the measurements of the PC kinetic by using the photons with energies $\hbar\omega = 1.608$ eV are sensitive to the quality of the surface and may be used to test methods of preparation of detector plates.

The PC kinetics measured for $\hbar\omega > E_g$ and $\hbar\omega < E_g$ indicated a minor influence of holes on the observed phenomena.

We described our results with three models: Hecht [2], Hecht-Many [3], and Ridzonova *et al.* [13]. The best agreement was obtained for the model by Ridzonova *et al*. The model considers the non-uniform distribution of the electric field.

We also point out that when the metallic contacts (Au, Pt) are deposited directly onto the surface of a high-resistive (Cd,Mn)Te sample with orientation (111), to obtain high resistivities, it is better to place the anode on the A (cadmium) side of the plate. However, that observation did not depend on the sample preparation, i.e., the concentration of HCl used for the final etching.

D.M.K., A.W.; supervision, A.M., J.I.-H., M.M. All authors have read and agreed to the published version of the manuscript.

**Funding:** This research was funded by the Polish National Centre for Research and Development, grant number TECHMATSTRATEG1/346720/8/NCBR/2017, and by the Foundation for Polish Science through the IRA Programme co-financed by EU within SG OP, grant number MAB/2017/1.

**Institutional Review Board Statement:** Not applicable.

**Informed Consent Statement:** Not applicable.

**Data Availability Statement:** The data presented in this study are available on request from the corresponding author.

**Acknowledgments:** The authors would like to gratefully acknowledge Sirs Marek Zubrzycki, Stanisław Jabłoński, Janusz Gdański, Adam Marciniak, Michał Kochański and Marcin Dopierała for their technical support during experiments. The authors would like to warmly thank Professor Roman Grill from Charles University in Prague (Czech Republic) for the insightful discussion and valuable comments on the manuscript.

**Conflicts of Interest:** The authors declare no conflict of interest. The funders had no role in the design of the study; in the collection, analyses, or interpretation of data; in the writing of the manuscript, or in the decision to publish the results.

**References**

1. Mycielski, A.; Wardak, A.; Kochanowska, D.; Witkowska-Baran, M.; Szot, M.; Jakieła, R.; Domagała, J.Z.; Kowalczyk, L.; Kochański, M.; Janusz, G.; et al. CdTe-based crystals with Mg, Se, or Mn as materials for X and gamma ray detectors: Selected physical properties. *Prog. Cryst. Growth Charact. Mater.* **2021**, *67*, 100543, doi:10.1016/j.pcrysgrow.2021.100543.

2. Hecht, K. Zum Mechanismus des lichtelektrischen Primärstromes in isolierenden Kristallen. *Zeitschrift für Phys.* **1932**, *77*, 235–245.

3. Many, A. High-field effects in photoconducting cadmium sulphide. *J. Phys. Chem. Solids* **1965**, *26*, 575–585, doi:https://doi.org/10.1016/0022-3697(65)90133-2.

4. Cui, Y.; Wright, G.W.; Ma, X.; Chattopadhyay, K.; James, R.B.; Burger, A. DC photoconductivity study of semi-insulating Cd1-xZnxTe crystals. *J. Electron. Mater.* **2001**, *30*, 774–778, doi:10.1007/BF02665871.

5. Cui, Y.; Groza, M.; Hillman, D.; Burger, A.; James, R.B. Study of surface recombination velocity of Cd1-

xZnxTe radiation detectors by direct current photoconductivity. *J. Appl. Phys.* **2002**, *92*, 2556–2560, doi:10.1063/1.1497696.

6. Levi, A.; Schieber, M.M.; Burshtein, Z. Carrier surface recombination in HgI2 photon detectors. *J. Appl. Phys.* **1983**, *54*, 2472–2476, doi:10.1063/1.332363.

7. Levi, A.; Burger, A.; Nissenbaum, J.; Schieber, M.; Burshtein, Z. Search for improved surface treatment procedures in fabrication of HgI2 X-ray spectrometers. *Nucl. Instruments Methods* **1983**, *213*, 35–38, doi:https://doi.org/10.1016/0167-5087(83)90039-X.

8. Burshtein, Z.; Akujieze, J.K.; Silberman, E. Carrier surface generation and recombination effects in photoconduction of HgI2 single crystals. *J. Appl. Phys.* **1986**, *60*, 3182–3187, doi:10.1063/1.337733.

9. Zappettini, A.; Zha, M.; Marchini, L.; Calestani, D.; Mosca, R.; Gombia, E.; Zanotti, L.; Zanichelli, M.; Pavesi, M.; Auricchio, N.; et al. Boron oxide encapsulated vertical Bridgman grown CdZnTe crystals as X-ray detector material. In Proceedings of the 2008 IEEE Nuclear Science Symposium Conference Record; 2008; pp. 118–121.

10. Zanichelli, M.; Pavesi, M.; Zappettini, A.; Marchini, L.; Auricchio, N.; Caroli, E.; Manfredi, M. Characterization of bulk and surface transport mechanisms by means of the photocurrent technique. *IEEE Trans. Nucl. Sci.* **2009**, *56*, 3591–3596, doi:10.1109/TNS.2009.2032098.

11. Zanichelli, M.; Santi, A.; Pavesi, M.; Zappettini, A. Charge collection in semi-insulator radiation detectors in the presence of a linear decreasing electric field. *J. Phys. D. Appl. Phys.* **2013**, *46*, 365103, doi:10.1088/0022-3727/46/36/365103.

12. Ling, Y.; Min, J.; Liang, X.; Zhang, J.; Yang, L.; Zhang, Y.; Li, M.; Liu, Z.; Wang, L. Carrier transport performance of Cd0.9Zn0.1Te detector by direct current photoconductive technology. *J. Appl. Phys.* **2017**, *121*, 034502, doi:10.1063/1.4974201.

13. Ridzonova, K.; Belas, E.; Grill, R.; Pekarek, J.; Praus, P. Space-charge-limited photocurrents and transient currents in (Cd,Zn)Te radiation detectors. *Phys. Rev. Appl.* **2020**, *13*, 064054, doi:10.1103/PhysRevApplied.13.064054.

14. Shen, M.; Zhang, J.; Wang, L.; Min, J.; Wang, L.; Liang, X.; Huang, J.; Tang, K.; Liang, W.; Meng, H. Investigation on the surface treatments of CdMnTe single crystals. *Mater. Sci. Semicond. Process.* **2015**, *31*, 536–542, doi:10.1016/j.mssp.2014.12.051.
20

15. Fiederle, M.; Eiche, C.; Salk, M.; Schwarz, R.; Benz, K.W.; Stadler, W.; Hofmann, D.M.; Meyer, B.K. Modified compensation model of CdTe. *J. Appl. Phys.* **1998**, *84*, 6689–6692, doi:10.1063/1.368874.

16. Chu, M.; Terterian, S.; Ting, D.; Wang, C.C.; Gurgenian, H.K.; Mesropian, S. Tellurium antisites in CdZnTe. *Appl. Phys. Lett.* **2001**, *79*, 2728–2730, doi:10.1063/1.1412588.

17. Brown, P.D.; Durose, K.; Russell, G.J.; Woods, J. The absolute determination of CdTe crystal polarity. *J. Cryst. Growth* **1990**, *101*, 211–215, doi:https://doi.org/10.1016/0022-0248(90)90968-Q.

18. Horodyský, P.; Hlídek, P. Free-exciton absorption in bulk CdTe: Temperature dependence. *Phys. Status Solidi B* **2006**, *243*, 494–501, doi:10.1002/pssb.200541402.

19. Horodyský, P.; Grill, R.; Hlídek, P. Band-edge photoluminescence in CdTe. *Phys. Status Solidi B* **2006**, *243*, 2882–2891, doi:10.1002/pssb.200642272.

20. Frenkel, J. On pre-breakdown phenomena in insulators and electronic semi-conductors. *Phys. Rev.* **1938**, *54*, 647–648, doi:10.1103/PhysRev.54.647.

21. Hartke, J.L. The Three-Dimensional Poole-Frenkel Effect. *J. Appl. Phys.* **1968**, *39*, 4871–4873, doi:https://doi.org/10.1063/1.1655871.

22. Dussel, G.A.; Bube, R.H. Electric field effects in trapping processes. *J. Appl. Phys.* **1966**, *37*, 2797–2804, doi:10.1063/1.1782126.

23. Martini, M.; McMath, T.A. Trapping and detrapping effects in lithium-drifted germanium and silicon detectors. *Nucl. Instruments Methods* **1970**, *79*, 259–276, doi:https://doi.org/10.1016/0029-554X(70)90149-7.

24. Bale, D.S.; Szeles, C. Nature of polarization in wide-bandgap semiconductor detectors under high-flux irradiation: Application to semi-insulating $Cd_{1-x}Zn_xTe$. *Phys. Rev. B* **2008**, *77*, 035205, doi:10.1103/PhysRevB.77.035205.

25. Franc, J.; Dědič, V.; Sellin, P.J.; Grill, R.; Veeramani, P. Radiation induced control of electric field in Au/CdTe/In structures. *Appl. Phys. Lett.* **2011**, *98*, 232115, doi:10.1063/1.3598414.

26. Wardak, A.; Szot, M.; Witkowska-Baran, M.; Avdonin, A.; Kochanowska, D.; Łusakowska, E.; Mycielski, A. Internal electric field in (Cd,Mn)Te and (Cd,Mg)Te studied by the Pockels effect. *J. Cryst. Growth* **2019**, *526*, 125217, doi:https://doi.org/10.1016/j.jcrysgro.2019.125217.

27. Owens, A.; Peacock, A. Compound semiconductor radiation detectors. *Nucl. Instruments Methods Phys. Res.*

*A* **2004**, *531*, 18–37, doi:doi:10.1016/j.nima.2004.05.071.

28. Najam, L.A.; Jamil, N.Y.; Yousif, R.M. Comparison in Mobility, Transit Time and Quality Factor Between CdMnTe and CdZnTe Detectors. *African Rev. Phys.* **2012**, *7*, 269–272.

29. Rafiei, R.; Reinhard, M.I.; Kim, K.; Prokopovich, D.A.; Boardman, D.; Sarbutt, A.; Watt, G.C.; Bolotnikov, A.E.; Bignell, L.J.; James, R.B. High-purity CdMnTe radiation detectors: A high-resolution spectroscopic evaluation. *IEEE Trans. Nucl. Sci.* **2013**, *60*, 1450–1456, doi:10.1109/TNS.2013.2243167.